\documentclass[english,floatfix,twocolumn,showkeys,superscriptaddress,nofootinbib,aps,prd]{revtex4}
\usepackage[latin1]{inputenc}
\usepackage[T1]{fontenc}
\usepackage{lmodern}
\usepackage{amsmath}
\usepackage{amssymb}
\usepackage{graphicx}
\usepackage{esint}
\usepackage{longtable}
\usepackage{dcolumn}
\usepackage{babel} 
\usepackage{csquotes} 
\usepackage{color} 
\usepackage{textcomp}

\begin{document}

\title{Bianchi type I cosmology with a Kalb--Ramond background field}

\author{R. V. Maluf}
\email{r.v.maluf@fisica.ufc.br}
\affiliation{Universidade Federal do Ceará (UFC), Departamento de Física,\\ Campus do Pici, Fortaleza - CE, C.P. 6030, 60455-760 - Brazil}
\author{Juliano C. S. Neves}
\email{juliano.neves@unifal-mg.edu.br}
\affiliation{Instituto de Ciência e Tecnologia, Universidade Federal de Alfenas, \\ Rodovia José Aurélio Vilela,
11999, CEP 37715-400 Poços de Caldas, MG, Brazil}

\begin{abstract}
An effect of the Lorentz symmetry breaking is pointed out in the cosmological context. Using a Bianchi I geometry
coupled to the Kalb--Ramond field, a consequence of the Lorentz symmetry violation is indicated by a 
different rate of expansion in a given spatial direction. This article focuses on the coupling 
constant $\xi_1$, which generates, from the Kalb--Ramond field, all three coefficients that give 
rise to the Lorentz violation in the gravity sector of the minimal Standard Model Extension.
The coupling constant $\xi_1$ increases the rate of expansion of the universe in a given direction 
during a dark energy era. As a consequence, a range of validity of that coupling constant is also obtained.  
 
\end{abstract}


\maketitle

\section{introduction}

There is a common statement that both the Lorentz and the $CPT$ symmetries violations would be
 fingerprints of a brand new physics at the Planck scale. 
It has been shown how such violations occur in different contexts, including 
string theory \cite{Kostelecky:1988zi,Kostelecky:1991ak}, noncommutative 
field theory \cite{Carroll:2001ws}, quantum loop gravity \cite{Gambini:1998it,Bojowald:2004bb}, 
and in the extra dimensions alternatives \cite{Rizzo:2005um,Santos:2012if}. 
A theoretical framework that includes a lot of possible Lorentz-violating effects and, at the same time, 
allows us to calculate modifications of quantities (that would be measured in experiments) is 
the Standard-Model Extension (SME) \cite{Colladay:1996iz,Colladay:1998fq}. 
The SME is an effective field theory that is in agreement with the gauge structure of the Standard Model (SM) 
of particle physics, 
allowing then the inclusion of Lorentz-violating coefficients that break the particle Lorentz invariance, 
but such coefficients still preserve the invariance under observer 
Lorentz transformations \cite{Kostelecky:2000mm}. 
The gravitational sector of the SME was introduced in Ref. \cite{Kostelecky:2003fs} and provides modifications 
on the Einstein--Hilbert term, including dynamic background fields coupled with both curvature and torsion. 

Since its formulation in the late 1990s  by Colladay and Kostelecký  \cite{Colladay:1996iz,Colladay:1998fq}, 
the SME has been 
conceived of as background for a variety of investigations like neutral meson 
oscillations \cite{Kostelecky:2001ff}, neutrino oscillations \cite{Diaz:2011ia}, 
clock-comparison tests \cite{Kostelecky:2018fmc}, spin-polarized torsion pendulum \cite{Bluhm:1999ev}, 
penning-trap tests of quantum electrodynamics \cite{Ding:2016lwt}, 
hydrogen and antihydrogen spectroscopy \cite{Bluhm:1998rk}, 
muon decay \cite{Noordmans:2014hxa}, and optical and microwave cavities \cite{Mewes:2012sm}. 
A large number of stringent constraints on Lorentz-violating coefficients were obtained in many sectors of 
the SME, especially for particles like photons, neutrinos, and leptons.\footnote{A compilation of several 
bounds can be found in Ref. \cite{Kostelecky:2008ts}.}

However, in contrast to the flat case, curved spacetimes do not support explicit 
Lorentz symmetry breaking (with fixed and nondynamical Lorentz-violating backgrounds) due to 
inconsistencies with both the Bianchi identities and conservation laws \cite{Kostelecky:2003fs}. 
A mechanism capable of overcoming these difficulties is the spontaneous 
Lorentz symmetry breaking \cite{Moffat:2002nu}, which works similarly to the Higgs mechanism, 
adding then a potential term to the Lagrange density constructed in such a way that its minimum lies 
at a nonzero vacuum expectation value (VEV) for either the background vector field or the background tensor 
field \cite{Bluhm:2007bd,Maluf:2013nva,Maluf:2014dpa} (whether vector or tensor fields, such background
fields are also called bumblebee fields). 
In this context, studying the behavior of gravity in the strong-field regime with the presence of 
spontaneous Lorentz violation is a tough task that has attracted attention in recent years, and several works involving 
solutions of compact objects like black holes \cite{Casana:2017jkc,Lessa:2019bgi,Ding:2020kfr,Maluf:2020kgf} 
and topologically nontrivial structures like wormholes \cite{Ovgun:2018xys,Lessa:2020imi} have been carried out. 
In the cosmological scenario, effects of the  Lorentz violation on the 
Friedmann-Lemaître-Robertson-Walker (FLRW) universe were studied in Refs. \cite{Bertolami:2005bh,Nilsson:2018ocp,Capelo:2015ipa}, 
and recently the impact of the Lorentz symmetry breaking on anisotropies of the cosmic microwave background
radiation was examined by 
us in Ref. \cite{Maluf:2021lwh}, in which the Bianchi type I geometry was assumed. 
In the latter work, we adopted a vector field in order to break the Lorentz
symmetry. Here we will continue these investigations on Lorentz symmetry breaking, assuming an antisymmetric 2-tensor background field coupled to gravity in the Bianchi I geometry instead of a  background vector  field.
 
As we said, the geometry adopted here is the Bianchi type I, which generalizes the flat FLRW metric from the
standard cosmology. The Bianchi I cosmology takes into consideration three independent scale factors, one for
each spatial direction, and is able to disagree with the cosmological principle, promoting then an anisotropic universe.\footnote{See works on Bianchi I cosmology in several contexts \cite{Maluf:2021lwh,Russell,Alexeyev:2000eb,Saha:2006iu,Rao:2008zzd,Rikhvitsky:2011js,Carloni:2013hna}.} 
In our previous work  \cite{Maluf:2021lwh}, the difference of a scale factor
in a given direction (compared to others) comes from the bumblebee vector field. On the other hand, now 
we will adopt an antisymmetric 2-tensor background field, known in the literature as the 
Kalb--Ramond field \cite{Kalb:1974yc}, studied in other contexts in Refs. \cite{Elizalde:2018rmz,Elizalde:2018now}.
 It is worth emphasizing
that a tensor field turns our task into an even more tough task. 
As we will see, for this case the background tensor field---by means of
its equation of motion---give us a more stringent constraint compared to the background vector field case, although such 
a constraint is ruled out by recent observations.  

As we mentioned before, the Kalb--Ramond field was adopted in several contexts. In cosmology,
the context of this article, the authors of Refs. \cite{Elizalde:2018rmz,Elizalde:2018now}
 built solutions for the inflationary period in the $F(R)$ gravity
with the Kalb--Ramond tensor field. They were able to show that in the early universe such an antisymmetry tensor
field could be important in the inflationary period. On the other hand, Ref. \cite{Aashish:2018lhv} 
adopted both the Einsteinian gravity
and the nonminimal model for the Kalb--Ramond field (like us) in order to obtain a de Sitter 
phase and the inflationary period.
But such works, in the cosmological context, chose the FLRW metric as the geometry of the universe, 
not the Bianchi I like our work, and do not assume the Lorentz symmetry breaking. 
Moreover, our focus is not on the early universe, our goals are the late universe 
(driven by dark energy) and a measurable effect due to the presence of a nonzero VEV for the Kalb--Ramond 
field, which breaks spontaneously the local Lorentz symmetry. 
 
Regarding observations, here we intend to show the importance of a coupling constant from the SME
on the spacetime dynamics. That is our main motivation. According to the SME \cite{Bailey:2006fd}, 
a specific parameter of the model (indicated as $\xi_1$) generates all three Lorentz-violating coefficient fields, 
namely the fields $u$, $s^{\mu\nu}$, and $t^{\kappa\lambda\mu\nu}$. In particular, the $t$-coefficient is written 
in terms of the background field or Kalb--Ramond field when it is contracted with the Riemann tensor from 
the $\xi_{1}$ coupling \cite{Altschul:2009ae}. From the phenomenological perspective,
the role played by the $t$-coefficient is an interesting and 
few studied issue, since previous works like Refs. \cite{Lessa:2019bgi,Lessa:2020imi} did not take into 
account the  coupling constant $\xi_1$.\footnote{Recently, an exception is made in Ref. \cite{Maluf:2021ywn},
 in which one of us studied that coefficient using a wormhole geometry.}
 According to Ref. \cite{Altschul:2009ae,Bonder:2015maa,Nilsson:2018ocp},
 the $t$-coefficient gives rise to the so-called $t$-puzzle. Interestingly, such a coefficient produces no contribution
 to the gravitational field equations in the linearized case, and that is the origin for the aforementioned puzzle. 
 However, as will see, here the gravitational
 field equations will not be linearized, thus the contributions of all three Lorentz-violating coefficients are still present. 
 In a cosmological context,
 the role played by the $t$-coefficient was studied in the FLRW spacetime in Ref. \cite{Nilsson:2018ocp}. 
 But as we said, in this article we adopt
 the Bianchi I geometry.    
 As we will see, the coupling constant $\xi_1$, which is responsible 
 for producing a nonzero VEV for the $t^{\kappa\lambda\mu\nu}$ field (breaking then the Lorentz symmetry), 
 increases the cosmic acceleration in a given direction during a dark energy era.

The article is structured as follows: in Sec. II the modified gravitational field equations are shown in the SME context, and
all three Lorentz-violating coefficients are defined. 
In Sec. III the Bianchi I geometry is presented, and the Kalb--Ramond field 
is coupled to it. The Friedmann-like equations are obtained, and then the solution for a dark energy dominated
universe is achieved in Sec. IV. The final remarks are in Sec. V.  

\section{The modified gravitational sector}

According to Ref. \cite{Altschul:2009ae}, at leading order in the curvature, the most general action for the (bumblebee) Kalb--Ramond model is given by
\begin{align}
 S  &=  \int d^{4}x\sqrt{-g}\bigg[\frac{1}{2\kappa}R-\frac{1}{12}H_{\mu\nu\lambda}H^{\mu\nu\lambda}-V+\mathcal{L}_{M}  +\frac{1}{2\kappa} \nonumber \\
 & \times \left(\xi_{1}B^{\kappa\lambda}B^{\mu\nu}R_{\kappa\lambda\mu\nu}+\xi_{2}B^{\lambda\nu}B_{\ \nu}^{\mu}R_{\lambda\mu}+\xi_{3}B^{\mu\nu}B_{\mu\nu}R\right)\bigg],
\label{Action}
\end{align}
where $\kappa=8\pi G_{N}/c^4$, such that $G_{N}$ is the Newtonian gravitational
constant, and $c$ is the speed of light in vacuum. The Kalb--Ramond field, indicated by $B_{\mu\nu}$, 
is an antisymmetric
2-tensor in the four-dimensional Riemannian manifold with its associated
field-strength tensor $H_{\mu\nu\lambda}$ defined by
\begin{equation}
H_{\mu\nu\lambda}=\partial_{\mu}B_{\nu\lambda}+\partial_{\lambda}B_{\mu\nu}+\partial_{\nu}B_{\mu\lambda},
\label{Field_Strength}
\end{equation}
with $H_{\mu\nu\lambda}$ being invariant under the gauge transformation
 $B_{\mu\nu}\rightarrow B_{\mu\nu}+\partial_{\mu}\Lambda_{\nu}-\partial_{\nu}\Lambda_{\mu}$.
Also, the potential $V$ is responsible for triggering spontaneously the
Lorentz violation inducing then a nonzero VEV for the bumblebee field, that is to say, 
\begin{equation}
\left\langle B_{\mu\nu} \right\rangle = b_{\mu\nu}.
\label{VEV}
\end{equation}
In the same way, the potential induces a nonzero VEV for the dual tensor field $\mathfrak{B}_{\mu\nu}$, indicated as
\begin{equation}
\left\langle \mathfrak{B}_{\mu\nu} \right\rangle = \left\langle \frac{1}{2}\epsilon_{\kappa\lambda\mu\nu} B^{\kappa\lambda} \right\rangle=\frac{1}{2}\epsilon_{\kappa\lambda\mu\nu} b^{\kappa\lambda},
\label{VEV2}
\end{equation}
where $\epsilon_{\kappa\lambda\mu\nu}$ is the totally antisymmetric Levi-Civita tensor. 
In general, the potential $V$ depends on $B_{\mu\nu}$ through the 
$B_{\mu\nu}B^{\mu\nu}$ and $B_{\mu\nu}\mathfrak{B}^{\mu\nu}$ scalars. 
However, as we will see in this work, the explicit form of the potential will not be relevant in our analysis.

Following the final terms of Eq. (\ref{Action}),  $\mathcal{L}_{M}$ regards the matter (or energy) content that will be 
specified later, the coupling constants $\xi_{1}$,
$\xi_{2}$, and $\xi_{3}$ represent all nonminimal nonderivative gravitational
couplings to the field $B_{\mu\nu}$ (coupling constants that are linear in the curvature, with mass dimension $[\xi]=M^{-2}$ in natural units), 
and $R$, $R_{\lambda\mu}$, and
$R_{\kappa\lambda \mu\nu}$ are the scalar curvature, the Ricci tensor and the Riemann tensor, respectively.

On the other hand, the minimal gravitational sector of the SME can
be written as 
\begin{equation}
S_{\mbox{LV}}^{(\mbox{min})}=\int d^{4}x\sqrt{-g}\frac{1}{2\kappa}\Big(-uR+s^{\mu\nu}R_{\mu\nu}^{T}+t^{\kappa\lambda\mu\nu}C_{\kappa\lambda\mu\nu}\Big),
\end{equation}
where $R_{\mu\nu}^{T}$ is the trace-free Ricci tensor, and $C_{\kappa\lambda\mu\nu}$
is the Weyl conformal tensor. In order to satisfy both the geometric constraints
of Riemann-Cartan manifolds and the conservation laws of general
relativity, the coefficients that account for the Lorentz violation, namely $u$, $s^{\mu\nu}$
and $t^{\kappa\lambda\mu\nu}$, must be functions of spacetime
position, and their underlying dynamics will depend on the choice
of the specific model. In particular, the theory described by action
(\ref{Action}) is capable of generating the necessary dynamics for
all three SME coefficients, and the correspondence with the Kalb--Ramond
field $B_{\mu\nu}$ can be read in the following form:
\begin{align}
u_{B}& =-(\frac{1}{6}\xi_{1}+\frac{1}{4}\xi_{2}+\xi_{3})B^{\alpha\beta}B_{\alpha\beta}, \\
s_{B}^{\mu\nu}& =(2\xi_{1}+\xi_{2})(B_{\ \alpha}^{\mu}B^{\nu\alpha}-\frac{1}{4}g^{\mu\nu}B^{\alpha\beta}B_{\alpha\beta}), 
\end{align}
\begin{align}
t_{B}^{\kappa\lambda\mu\nu} & =\frac{2}{3}\xi_{1}(B^{\kappa\lambda}B^{\mu\nu}+\frac{1}{2}B^{\kappa\mu}B^{\lambda\nu}-\frac{1}{2}B^{\kappa\nu}B^{\lambda\mu})\nonumber \\
 & -\frac{1}{2}\xi_{1}(g^{\kappa\mu}B_{\ \alpha}^{\lambda}B^{\nu\alpha}-g^{\lambda\mu}B_{\ \alpha}^{\kappa}B^{\nu\alpha}\nonumber \\
 & -g^{\kappa\nu}B_{\ \alpha}^{\lambda}B^{\mu\alpha}+g^{\lambda\nu}B_{\ \alpha}^{\kappa}B^{\mu\alpha})\nonumber \\
 & +\frac{1}{6}\xi_{1}(g^{\kappa\mu}g^{\lambda\nu}-g^{\lambda\mu}g^{\kappa\nu})B^{\alpha\beta}B_{\alpha\beta}.
\end{align}

It is worth emphasizing that the spontaneous Lorentz-breaking effects controlled by the fields or coefficients
 $u$, $s^{\mu\nu}$ and $t^{\kappa\lambda\mu\nu}$ have been widely
explored topics. In particular, black hole solutions and anisotropies
in the cosmic microwave background radiation were recently investigated by us in
Refs. \cite{Maluf:2020kgf,Maluf:2021lwh}. However, 
the description of that effects depends on
the fields that are chosen to compose the mentioned coefficients. In particular, the physics and consequently the
phenomenology
of the $t$-coefficient can not be achieved by vector fields (adopted by us in the mentioned works), it requires
the presence of an antisymmetric 2-tensor, being the least explored
in the literature so far. Thus, as we said in Introduction, our main focus will be on the coupling constant $\xi_{1}$,
which is responsible for producing a nonzero VEV for the field $t_B^{\kappa\lambda\mu\nu}$.

Varying the action (\ref{Action}) with respect to the metric tensor, 
one obtains  the modified gravitational field equations: 
\begin{equation}
G^{\mu\nu}  =\kappa T_{M}^{\mu\nu}+\kappa T_{B} ^{\mu\nu} + T_{\xi_{1}}^{\mu\nu}+ T_{\xi_{2}}^{\mu\nu}+ T_{\xi_{3}}^{\mu\nu}. 
 \label{Field_equations}
\end{equation}
The left side of (\ref{Field_equations}) is the usual Einstein
tensor $G_{\mu\nu}=R_{\mu\nu}-\frac{1}{2}Rg_{\mu\nu}$, as  the terms on the
right side represent  energy-momentum tensors whose sources are both the
matter content $T_{M}^{\mu\nu}$ and the Kalb--Ramond field $B_{\mu\nu}$. 
Contributions in Eq. (\ref{Field_equations}) 
due to the Kalb--Ramond field come from the kinetic and potential
terms in $T_{B}^{\mu\nu}$ and the three couplings, $T_{\xi_{1}}^{\mu\nu}$,
 $T_{\xi_{2}}^{\mu\nu}$, and $T_{\xi_{3}}^{\mu\nu}$. For
$B$-terms, we have explicitly
\begin{align}
T_{B}^{\mu\nu} & =\frac{1}{2}H^{\alpha\beta\mu}H_{\ \alpha\beta}^{\nu}-\frac{1}{12}g^{\mu\nu}H^{\alpha\beta\gamma}H_{\alpha\beta\gamma}\nonumber \\
 & -g^{\mu\nu}V+4B^{\alpha\mu}B_{\alpha}^{\ \nu}V',
\end{align}
in which the operator $(')$ stands for derivative with respect to the potential
argument, and for the sake of simplicity the dependence of the potential
$V$ with respect to $B_{\mu\nu}$ is assumed in the form
\begin{equation}
V\equiv V(B_{\mu\nu}B^{\mu\nu}-x),
\end{equation}
with $x$ playing the role of a real number, meaning the VEV of the following invariant
\begin{equation}
x\equiv\left\langle B_{\mu\nu}B^{\mu\nu}\right\rangle =\left\langle g^{\alpha\mu}\right\rangle \left\langle g^{\beta\nu}\right\rangle b_{\alpha\beta}b_{\mu\nu}.
\end{equation}
Note that $\left\langle g^{\mu\nu}\right\rangle $ is the
VEV of the inverse metric. Since we are interested in investigating the Lorentz-violating  
effects due to a nonzero VEV for the $B_{\mu\nu}$ field, let us restrict ourselves to the case where 
the primary fields are frozen in their vacuum values, i.e.,
\begin{equation}
B_{\mu\nu}=b_{\mu\nu},\ \ \ \ g_{\mu\nu}=\left\langle g_{\mu\nu}\right\rangle ,
\end{equation}
breaking then both the local Lorentz and diffeomorphism symmetries. 
Also, the extreme conditions $V=V'=0$, which fix the vacuum state, are assumed and satisfied here.

The remaining terms in Eq. (\ref{Field_equations}) originated from the nonminimal gravitational couplings are
explicitly written as
\begin{align}
T_{\xi_{1}}^{\mu\nu} & =\xi_{1}\left(\frac{1}{2}g^{\mu\nu}B^{\alpha\beta}B^{\gamma\delta}R_{\alpha\beta\gamma\delta}+\frac{3}{2}B^{\beta\gamma}B^{\alpha\mu}R_{\ \alpha\beta\gamma}^{\nu}\right.\nonumber \\
 & +\frac{3}{2}B^{\beta\gamma}B^{\alpha\nu}R_{\ \alpha\beta\gamma}^{\mu}+\nabla_{\alpha}\nabla_{\beta}B^{\alpha\mu}B^{\nu\beta}\nonumber \\
 & +\left.\nabla_{\alpha}\nabla_{\beta}B^{\alpha\nu}B^{\mu\beta}\right),
\end{align}

\begin{align}
T_{\xi_{2}}^{\mu\nu} & =\xi_{2}\left(\frac{1}{2}g^{\mu\nu}B^{\alpha\gamma}B_{\ \gamma}^{\beta}R_{\alpha\beta}-B^{\alpha\mu}B^{\beta\nu}R_{\alpha\beta}\right.\nonumber \\
 & -B^{\alpha\beta}B_{\ \beta}^{\mu}R_{\ \alpha}^{\nu}-B^{\alpha\beta}B_{\ \beta}^{\nu}R_{\ \alpha}^{\mu}\nonumber \\
 & +\frac{1}{2}\nabla_{\alpha}\nabla^{\mu}B_{\ \beta}^{\nu}B^{\alpha\beta}+\frac{1}{2}\nabla_{\alpha}\nabla^{\nu}B_{\ \beta}^{\mu}B^{\alpha\beta}\nonumber \\
 & -\left.\frac{1}{2}\nabla^{2}B^{\alpha\mu}B_{\alpha}^{\ \nu}-\frac{1}{2}g^{\mu\nu}\nabla_{\alpha}\nabla_{\beta}B^{\alpha\gamma}B_{\ \gamma}^{\beta}\right),
\end{align}
and 
\begin{align}
T_{\xi_{3}}^{\mu\nu} & =\xi_{3}\left(\nabla^{\mu}\nabla^{\nu}B^{\alpha\beta}B_{\alpha\beta}-g^{\mu\nu}\nabla^{2}B^{\alpha\beta}B_{\alpha\beta}\right.\nonumber \\
 & -\left.B^{\alpha\beta}B_{\alpha\beta}G^{\mu\nu}+2B^{\alpha\mu}B_{\ \alpha}^{\nu}R\right).
\end{align}

Now varying the action (\ref{Action}) with respect to the antisymmetric tensor field $B_{\mu\nu}$, keeping the metric and matter fields fixed, gives us the field equations for the Kalb--Ramond field, namely
\begin{equation}
\nabla_{\alpha}H^{\alpha\mu\nu}=J_{V}^{\mu\nu}+J_{R}^{\mu\nu},
\label{KReom}
\end{equation}
with the potential current $J_{V}^{\mu\nu}$ written in the form
\begin{equation}
J_{V}^{\mu\nu}=4V'B^{\mu\nu},
\end{equation}
and the curvature current $J_{R}^{\mu\nu}$ being given by
\begin{equation}
J_{R}^{\mu\nu}=-\frac{2\xi_{1}}{\kappa}B_{\alpha\beta}R^{\alpha\beta\mu\nu}+\frac{\xi_{2}}{\kappa}B_{\alpha}^{\ [\mu}R^{\nu]\alpha}-\frac{2\xi_{3}}{\kappa}B^{\mu\nu}R.
\end{equation}
It is worth pointing out that here we are explicitly disregarding any coupling between the matter fields 
and the Kalb--Ramond field. Such a possibility (a particular coupling between matter and $B_{\mu\nu}$) 
may imply changes 
in the conservation of the conventional matter currents, something beyond our scope in this article.

The conservation of the total current is immediately achieved from
the choice of the kinetic term for the field $B_{\mu\nu}$, satisfying then the condition
\begin{equation}
\nabla_{\mu}\left(J_{V}^{\mu\nu}+J_{R}^{\mu\nu}\right)=0.
\label{eq:covcurrentcons}
\end{equation}
Alternatively, assuming that the energy-momentum tensor from the matter content is covariantly
conserved $(\nabla_{\mu}T^{\mu\nu}_{M}=0)$, alongside the Bianchi identity for the Einstein tensor, 
that is $\nabla_{\mu}G^{\mu\nu}=0$,
 then the components $T_{B}^{\mu\nu}$, $T_{\xi_{1}}^{\mu\nu}$,
$T_{\xi_{2}}^{\mu\nu}$, and $T_{\xi_{3}}^{\mu\nu}$ of the total
energy-momentum tensor also lead to the condition
\begin{equation}
\kappa\nabla_{\mu}T_{B}^{\mu\nu}=-\nabla_{\mu}\left(T_{\xi_{1}}^{\mu\nu}+T_{\xi_{2}}^{\mu\nu}+ T_{\xi_{3}}^{\mu\nu}\right).
\end{equation}

As we mentioned earlier, we are going to focus on the phenomenology of the 
$\xi_1$ coupling constant of the modified field equations. That means \textit{the couplings $\xi_2$ and $\xi_3$ are
set to zero from here}.

\section{The geometry and the background field}

\subsection{The Bianchi type I metric and the Kalb--Ramond field}

In order to study the influence of the coupling constant $\xi_1$ on the spacetime dynamics, we assume 
the Bianchi I geometry as the cosmological spacetime, allowing then that spatial directions are 
different from each other, illustrating (as we will see) an effect of the Lorentz symmetry violation on spacetime. 
The Bianchi I metric (or line element) in the Cartesian coordinates $(x_1,x_2,x_3)$ is written simply as
\begin{equation}
ds^{2}=-c^{2}dt^{2}+a_{1}(t)^{2}dx_{1}^{2}+a_{2}(t)^{2}dx_{2}^{2}+a_{3}(t)^{2}dx_{3}^{2},\label{eq:Bianchimetric}
\end{equation}
in which $a_1(t)$, $a_2(t)$, and $a_3(t)$ are scale factors for each spatial direction, $(x_1,x_2,x_3)$ respectively. 
Thus, the Hubble parameter is defined for each spatial direction, i.e., 
\begin{equation}
H_{i}=\frac{\dot{a}_{i}(t)}{a_{i}(t)},\ \text{with \textit{i}=1,2,3},
\label{Hubble_parameter}
\end{equation}
where dot means derivative with respect to the time coordinate
$t$ or the cosmic time. Therefore, in this context, contrary to the FLRW cosmology, we could  have 
different directional Hubble parameters.

Now let us set the vacuum value of the Kalb--Ramond field $B_{\mu\nu}$ in order to describe its influence on the
spacetime dynamics. 
In four dimensions, an antisymmetric 2-tensor like the Kalb--Ramond field
 can be decomposed into two independent 3-vectors like
\begin{equation}
b_{0i}=-\Sigma^{i}\ \ \mbox{and} \ \ b_{jk}=\epsilon_{jkl}\Xi^{l}.
\end{equation}
This is similar to the electric and magnetic decomposition of the field strength $F_{\mu\nu}$ in 
Maxwell electrodynamics. Here the 3-vector fields $\vec{\Sigma}$ and $\vec{\Xi}$ encode 
the six degrees of freedom of the vacuum value $b_{\mu\nu}$. However, it is always possible to 
perform local observer Lorentz transformations in order to rewrite $b_{\mu\nu}$ in a simple block-diagonal form. 
For our purpose, following Ref. \cite{Altschul:2009ae}, this simple and special form can be conceived of as
\begin{equation}
b_{\mu\nu}=\begin{pmatrix}0 & -\mathcal{A}a_{1} & 0 & 0\\
\mathcal{A} a_{1} & 0 & 0 & 0\\
0 & 0 & 0 & \mathcal{B}a_{2}a_{3}\\
0 & 0 & -\mathcal{B}a_{2}a_{3} & 0
\end{pmatrix},
\label{bfield}
\end{equation}
where $\mathcal{A}$ and $\mathcal{B}$ are real numbers (as we pointed out, both are interpreted 
as the \enquote{electric} and 
\enquote{magnetic} parts, respectively, of the bumblebee field). 
In this special frame $x=b_{\mu\nu}b^{\mu\nu}=-2(\frac{\mathcal{A}^{2}}{c^2}-\mathcal{B}^{2})$,
ensuring that $b_{\mu\nu}$ produces a constant invariant scalar. Besides, our analysis is simplified as 
we have now only two Lorentz-violating parameters $(\mathcal{A}$ and $\mathcal{B})$ instead of six initial
parameters that remain for a general case. 

Consequently, the field strength $H_{\mu\nu\lambda}$, 
defined in Eq. (\ref{Field_Strength}), is not identically zero according
to the field $b_{\mu\nu}$ chosen in Eq. (\ref{bfield}). There are six nonzero
components, namely
\begin{align}
H_{023}=& -H_{032}=H_{230}=-H_{203}=H_{302}=-H_{320} \nonumber \\
 =& \ \mathcal{B}(a_{3}\dot{a}_{2}+a_{2}\dot{a}_{3}).
\end{align}
As is clear from Eq. (\ref{KReom}), such components will be used in the calculations of the Kalb--Ramond field equations. 

\subsection{The energy-momentum tensor(s)}
In order to describe the matter-energy content, which will be a single component 
(excluding the background field), we assume the perfect fluid interpretation, which will allow us to describe 
the Kalb--Ramond field influence on the cosmic expansion.
Therefore, the matter-energy content will be indicated by the tensor
\begin{equation}
T^{\mu\nu}_M = \left[\rho_{M}(t) + \frac{p_{M}(t)}{c^2}  \right]u^{\mu}u^{\nu}+p_{M}(t)g^{\mu\nu},
\label{PF}
\end{equation}
in which $\rho_M (t)$ and $p_M(t)$ stand for the density and pressure, respectively, of the matter-energy
 content (as we will see the bumblebee field or the Kalb--Ramond field has its own energy-momentum tensor).
  The four-vector $u^{\mu}$ is the four-velocity of the fluid, and, in particular
 for the metric signature adopted here,  $u_{\mu}u^{\mu}=-c^2$. Like our previous work 
  \cite{Maluf:2021lwh},  the matter-energy content
 is an isotropic fluid, and, as we will see, anisotropies come from the Kalb--Ramond field.

On the other hand, the corresponding Kalb--Ramond energy-momentum tensor in the Bianchi I cosmology, 
from the specific form of that field, given by
Eq. (\ref{bfield}), reads
 \begin{equation}
 \left(T_B \right) _{\ \nu}^{\mu}= \left(\begin{array}{cccc}
-\rho_B c^2\\
 &p_1 \\
 &  & p_2 \\
 &  &  & p_3
\end{array}\right),
\label{B_Energy-momentum}
\end{equation}
with
\begin{equation}
-\rho_B c^2=-p_1=p_2=p_3=-\frac{\mathcal{B}^2}{2c^2}\left(H_2+H_3 \right)^2.
\label{Comp_B}
\end{equation}  
Thus, contrary to the energy-momentum tensor for the matter field, it is evident the anisotropic
feature of the $T^{\mu\nu}_B$ field. As we can straightforwardly read, pressure in the $x_1$ direction
is different from other directions. Later we will assume that 
the $x_2$ and $x_3$ directions are identical ones, then $H_2=H_3$.

For the coupling $\xi_1$, its respective energy-momentum tensor is written as
\begin{equation}
 \left(T_{\xi_1} \right) _{\ \nu}^{\mu}= \left(\begin{array}{cccc}
-\rho_{\xi_1} c^2\\
 & P_1 \\
 &  & P_2 \\
 &  &  & P_3
\end{array}\right),
\label{Xi_Energy-momentum}
\end{equation}
with
\begin{align}
-\rho_{\xi_1} c^2  =&  -\frac{2\xi_1}{c^2}\bigg \{ \Big[ H_1H_2+H_1H_3 - 2\left( \dot{H_1}+ H_1^2\right) \Big]\frac{\mathcal{A}^2}{c^2} \nonumber \\
& + \mathcal{B}^2H_2H_3\bigg \},
\end{align}
\begin{align}
 P_1 = & -\frac{2\xi_1}{c^2}\bigg \{ \Big[ \dot{H_2}+\dot{H_3} - 2\left( \dot{H_1}+ H_1^2\right) +\left(H_2+H_3 \right)^2 \Big] \nonumber \\
& \times \frac{\mathcal{A}^2}{c^2}  - \mathcal{B}^2H_2H_3\bigg \}, \\
P_2=&  -\frac{2\xi_1}{c^2}\bigg \{\Big[ \dot{H_1}+ H_1^2\Big ] \frac{\mathcal{A}^2}{c^2} 
 +  \Big[ \dot{H_3}+H_3(H_1+2H_2  \nonumber \\
 & +H_3) \Big] \mathcal{B}^2 \bigg \}, \\
P_3=&  -\frac{2\xi_1}{c^2}\bigg \{\Big[ \dot{H_1}+ H_1^2\Big] \frac{\mathcal{A}^2}{c^2} 
 +  \Big[ \dot{H_2}+H_2(H_1+H_2 \nonumber \\
 &+2H_3) \Big]  \mathcal{B}^2 \bigg \}.
\end{align}
In this case, all three pressures could be different. But as we said, due to $T^{\mu\nu}_B$ components, 
the $P_2$ and $P_3$ components will be conceived of as identical ones. 

\subsection{The Friedmann-like equations}
From the modified field equations (\ref{Field_equations}), the background field (\ref{bfield}) and  
the total energy-momentum tensor 
$(T_T^{\mu\nu}=T_M^{\mu\nu}+T_B^{\mu\nu}+T^{\mu\nu}_{\xi_1}/\kappa)$,
we are able to write the Friedmann-like equations in the Bianchi I cosmology in the context studied here. By doing several
calculations, we have the following components: 
\begin{align}
H_1H_2+H_1H_3+H_2H_3=& \  8\pi G \left(\rho_M+\rho_B+\bar{\rho}_{\xi_1} \right), \label{Eq00} \\
\dot{H_2}+\dot{H_3}+H_2^2+H_3^2+H_2H_3=& -\frac{8\pi G}{c^2}\left(p_M+p_1+\bar{P}_1 \right), \label{Eq11}\\
\dot{H_1}+\dot{H_3}+H_1^2+H_3^2+H_1H_3=& -\frac{8\pi G}{c^2}\left(p_M+p_2+\bar{P}_2 \right), \label{Eq22} \\
\dot{H_1}+\dot{H_2}+H_1^2+H_2^2+H_1H_2=& -\frac{8\pi G}{c^2}\left(p_M+p_3+\bar{P}_3 \right), \label{Eq33}
\end{align}
where  $\bar{\rho}_{\xi_1}=\rho_{\xi_1}/\kappa$ and $\bar{P}_i=P_i/ \kappa$, with $i=1,2,3$.

It is worth calculating the mean Hubble parameter, which is defined as
\begin{equation}
H=\frac{1}{3}\left(H_1+H_2+H_3 \right).
\end{equation}
Then by using Eqs. (\ref{Eq00})-(\ref{Eq33}), one has a Friedmann-like equation also for the mean Hubble
parameter, that is,
\begin{align}
\dot{H}+3H^2= & \ 4\pi G \bigg \{\rho_M+\frac{4}{3}\rho_B +\bar{\rho}_{\xi_1}  \nonumber \\
& -\frac{1}{c^2}\Big[ p_M+\frac{1}{3}\left(\bar{P}_1+\bar{P}_2+\bar{P}_3 \right) \Big] \bigg \},
\label{H}
\end{align}
which will be very useful later.

\subsection{Constraints from the background field equations}
The Kalb--Ramond field equations, that is to say Eq. (\ref{KReom}), provides, 
after adopting the Bianchi I metric and the tensor field (\ref{bfield}), the following two
constraints. It is worth emphasizing that Eq. (\ref{eq:covcurrentcons}), which shows the current conservation,
 is satisfied only when the two constraints are adopted. 

\subsubsection*{First constraint}
The first one comes from the $\mu =0$ and $\nu =1$ component (which is equal to the minus 
$\mu=1$ and $\nu =0$ component)
 of that equation of motion, which is written as
\begin{equation}
\frac{4\mathcal{A}\xi_1\ddot{a}_1}{\kappa c^4 a_1^2}=0.
\label{Constraint_1}
\end{equation}
The above constraint implies that either $\mathcal{A}=0$ or $\ddot{a}_1=0$. The latter is ruled out by well-known
reports  \cite{SupernovaSearchTeam:1998fmf,SupernovaCosmologyProject:1998vns,Planck:2015fie}, that is,
with $\ddot{a}_1=0$ one has a null deceleration parameter $q_1=-\ddot{a}_1a_1/\dot{a}_1^2=0$ in the
$x_1$ direction or, equivalently, absence of an accelerated expansion in that direction. It is worth pointing out
that such a constraint is more stringent than that one obtained as the bumblebee field is a vector 
field \cite{Maluf:2021lwh}, although
in the case presented here this stringent constraint is ruled out by observations. 
 Having said that, we then assume---from the 
constraint (\ref{Constraint_1})---that $\mathcal{A}=0$ in the next calculations, i.e., the \enquote{electric} part of
$b_{\mu\nu}$ is null.

\subsubsection*{Second constraint}
The second constraint from the Kalb--Ramond field equations (\ref{KReom}) is given by
\begin{equation}
\dot{H_2}+\dot{H_3}+H_1(H_2 + H_3)-\frac{4\xi_1}{\kappa}H_2H_3=0.
\label{Constraint_2}
\end{equation}
It comes from the $\mu=2$ and $\nu=3$ component (again, it is equal to the minus $\mu=3$ 
and $\nu=2$ component). The two constraints will be very useful in order to solve the Friedmann-like equations. 
Another useful simplification
comes from the fact that $p_2=p_3$, according to the Kalb--Ramond energy-momentum tensor 
indicated in Eq. (\ref{Comp_B}). Therefore, as we said, the directions $x_2$ and $x_3$ will be considered equivalent.
From that assumption, the constraint (\ref{Constraint_2}) and the $\mu = \nu=1$ component of the
Friedmann-like equations, given by Eq. (\ref{Eq11}),  lead to
\begin{equation}
H_j^2=\frac{\kappa^2c^2}{4\left(\kappa+\xi_1 \right)}\left(\rho_M c^2- p_M\right), \\\  \mbox{for} \\\  j=2,3.
\label{H_j}
\end{equation}
Note that the above result is obtained just for the $\mathcal{A}=0$ case. 

\subsection{Energy-momentum conservation}
For a reliable physical model, the conservation of the energy-momentum tensor should be satisfied. 
Here the energy-momentum conservation, namely $\nabla _\nu T_T^{\mu\nu}= \nabla_\nu (\kappa T_M^{\mu\nu}+ \kappa T_B^{\mu\nu}+T_{\xi_1}^{\mu\nu})=0$,
is valid by using the two constraints mentioned before. With the aid of (\ref{Constraint_1})-(\ref{Constraint_2}), 
then the energy-momentum conservation leads to the well-known equation, i.e., 
\begin{equation}
\dot{\rho}_M= -3H \left(\rho_M + \frac{p_M}{c^2} \right).
\label{Matter_equation}
\end{equation} 
As we can see, the Kalb--Ramond field does not couple with the matter fields. 

\section{Solving the Friedmann-like equations for a dark energy dominated universe}
In order to solve the complicated system of equations (\ref{Eq00})-(\ref{Eq33}), we follow our 
previous work \cite{Maluf:2021lwh} in which
the Bianchi I geometry is conceived of as a small deviation from the FLRW metric. Contrary to the quoted work,
we are dealing with a tensor field, instead of a vector field  like in the previous work. 
It is worth emphasizing that
we are working on the late universe, not on the early universe and the inflationary period with a Kalb--Ramond field like
Refs. \citep{Elizalde:2018rmz,Elizalde:2018now,Aashish:2018lhv}. Most importantly, the motivation here is not 
to assign the antisymmetric background field as source of the
dark energy, but to indicate the origin of anisotropies in the accelerated expansion phase by means of the
Kalb--Ramond field in 
the Lorentz symmetry breaking context.
And, as we will see, anisotropies of that type will be given by different directional Hubble parameters. 

From the Friedman-like equations (\ref{Eq00})-(\ref{Eq33}), and with the aid of the mean 
Hubble parameter (\ref{H}), we have 
\begin{align}
\dot{H}_1+3HH_1= \ & 4\pi G\left(1+\delta \right) \left(\rho_M- \frac{p_M}{c^2} \right),\label{H1} \\
\dot{H}_j+3HH_j= \ & 4\pi G \left(\rho_M- \frac{p_M}{c^2} \right) \label{Hj},
\end{align}
where the dimensionless parameter $\delta$ is defined as
\begin{equation}
\delta = 2 \left(\kappa+\xi_1 \right)  \mathcal{B}^2,
\label{delta}
\end{equation}
with $0 < \delta < 1$.  As we can see, the geometry studied here also can be conceived 
of as  a small deviation from the FLRW metric. With $\delta = 0$, the 
Friedmann equations in the FLRW context are restored. As we note in Eq. (\ref{delta}), setting $\xi_1=0$
does not eliminate $\delta$ or even an effect of the Lorentz symmetry breaking on the geometry.
 This is due to the  fact that there is still a contribution from the \enquote{magnetic} part of the Kalb--Ramond field, 
 the kinetic term in the action (\ref{Action}).

Eqs. (\ref{H1}) and (\ref{Hj}) provide
a simpler equation for the mean Hubble parameter, that is to say, adding up those components one has
\begin{equation}
\dot{H}+3H^2=4\pi G \left(1+\frac{\delta}{3} \right) \left(\rho_M- \frac{p_M}{c^2} \right).
\label{Mean_Hubble}
\end{equation}
Thus, from the equation for the mean Hubble parameter, solutions of (\ref{H1}) and (\ref{Hj}) are straightforwardly written as
 \begin{equation}
 H_1(t)=\frac{1}{\mu (t)}\bigg[ K_1 + \int^{t}  \mu (t') 4\pi G \left(1+\delta \right) \bigg(\rho_M -\frac{p_M}{c^2} \bigg) dt' \bigg],
 \label{H1_solution}
 \end{equation}
 for the component in the $x_1$ direction, and 
 \begin{equation}
 H_j(t)=\frac{1}{\mu (t)}\bigg[ K_j + \int^{t} \mu (t') 4\pi G\bigg(\rho_M -\frac{p_M}{c^2} \bigg) dt' \bigg],
 \label{Hj_solution}
 \end{equation}
in which, as we said, $j=2,3$ are the components in the $x_2$ and $x_3$ directions, respectively. 
 In the above equations, $K_i$ (with $i=1,2,3$) are integration constants, and the function $\mu (t)$ 
 is defined as
\begin{equation}
\mu (t)= \exp \left(\int^{t} 3H(s) ds\right).
\label{mu}
\end{equation} 
Like our previous work \cite{Maluf:2021lwh},
 the constants $K_i$ are set to zero. Therefore, the difference from each spatial
direction is  given just by the influence of the Kalb--Ramond field on the spacetime geometry. Without the Kalb--Ramond
field, in the general relativity realm, the difference from each directional Hubble parameter comes from 
those integration constants. Adopting the background field brings us a physical origin for the difference among each
directional Hubble parameter, without considering the integration constants $K_i$.

\begin{figure}
\begin{centering}
\includegraphics[trim=0.5cm 0.4cm 0.8cm 0cm, clip=true,scale=0.57]{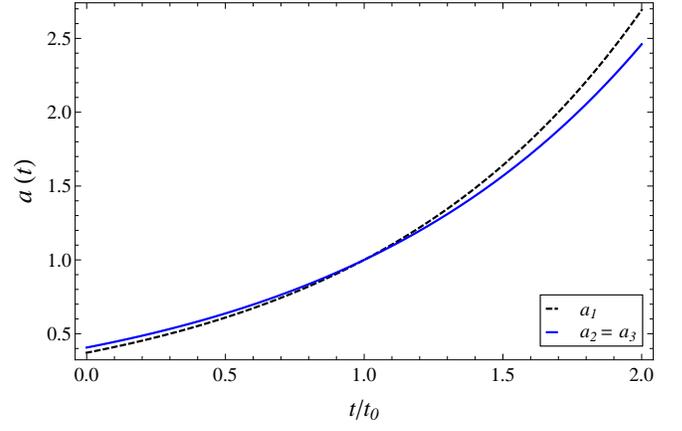}
\par\end{centering}
\caption{Scale factors (one for each spatial direction) for a dark energy dominated universe when the 
Kalb--Ramond assumes its VEV, 
breaking the Lorentz symmetry, in the Bianchi type I geometry. 
The parameter $t_0=1$ means the present time.}
\label{a_graphic}
\end{figure}

For a dark energy dominated universe, let us adopt the following equation of state for the cosmic fluid:
 $\omega_{DE}=p_{DE}/\rho_{DE}c^2=-1$ (which is equivalent to a vacuum spacetime, or $T^{\mu\nu}_M=0$,
 with a cosmological constant).  Consequently, from Eq. (\ref{Matter_equation}), 
 the dark energy density is constant, i.e., $\dot{\rho}_{DE}=0$. Then the mean Hubble parameter equation, indicated by 
 Eq. (\ref{Mean_Hubble}), give us as solution
\begin{equation}
H_{DE}=\left[\frac{8\pi G}{3} \left(1+\frac{\delta}{3} \right) \rho_{DE}\right]^{\frac{1}{2}}.
\end{equation}
With the mean Hubble parameter, in  this case for a dark energy dominated universe, we are able to 
calculate the directional Hubble parameters (\ref{H1_solution})-(\ref{Hj_solution}) with the aid of Eq. (\ref{mu}). 
After calculating them, one has
\begin{align}
H_1=& \left(1+\delta \right)\left[\frac{8\pi G}{\left(3+\delta \right)}\rho_{DE} \right]^{\frac{1}{2}}, \label{H1DE} \\
H_j = & \left[\frac{8\pi G}{\left(3+\delta \right)}\rho_{DE} \right]^{\frac{1}{2}} \label{HjDE}. 
\end{align}
For $\delta =0$, the constant Hubble parameter $H_0$ for a dark energy dominated universe in the FLRW perspective is
restored, i.e., $H_1=H_j=H_{DE}=H_0=\left(\frac{8\pi G \rho_{DE}}{3} \right)^\frac{1}{2}$. 
With the directional Hubble parameters, it is easy to get the corresponding scale factors, according to
the definition (\ref{Hubble_parameter}). Thus,
\begin{align}
a_1 (t) = & \exp^{H_1 \left(t-t_0 \right)}, \\
a_j (t) = & \exp^{H_j \left(t-t_0 \right)},
\end{align}
in which $t_0$ is the present time or the age of the universe. 
According to Fig. \ref{a_graphic}, the Kalb--Ramond field affects the $x_1$ direction accelerating it, showing then
that the $x_1$ direction has an expansion phase faster than other spatial directions. As this very difference comes from
the parameter $\delta$ and, consequently, from the coupling constant $\xi_1$, then we describe with that
scale factors the influence of that coupling constant on the spacetime dynamics, its phenomenology in this simplified
case.  

Lastly, the constraint (\ref{H_j}) and the above solution for $H_j$ in a dark energy dominated universe give us 
\begin{equation}
\delta = \frac{2\xi_1-\kappa}{\kappa}.
\label{delta2}
\end{equation}
Therefore, by using the definition of $\delta$, given by Eq. (\ref{delta}), 
we have the following value for the Lorentz-violating parameter $\mathcal{B}$ and, consequently,
 for the norm of the Kalb--Ramond or bumblebee field in the VEV, namely
\begin{equation}
\mathcal{B}^2=\frac{2\xi_1-\kappa}{2\kappa \left(\kappa+\xi_1 \right)} \ \ \mbox{and} \ \ b_{\mu\nu}b^{\mu\nu}=\frac{2\xi_1-\kappa}{\kappa \left(\kappa+\xi_1 \right)}.
\end{equation} 
On the other hand, assuming Eq. (\ref{delta}) and Eq. (\ref{delta2}), with $0 < \delta < 1$, that is, 
our cosmological solution is a small deviation from the FLRW
spacetime,  we have an interesting range of validity of the coupling constant $\xi_1$, i.e.,
\begin{equation}
\frac{\kappa}{2}< \xi_{1}<\kappa,
\end{equation}
producing then Lorentz symmetry breaking effects on a geometry that is close to the standard geometry in cosmology.

\section{Final remarks}
Lorentz symmetry violation has been a fruitful area in theoretical physics in the last decades. 
The ideia that violations of that symmetry can reveal hints of a quantum theory of gravity is an attracting topic.
Here we studied the spontaneous Lorentz violation caused by an antisymmetric  tensor field, the Kalb--Ramond field. 
In particular, we focused on
the  coupling constant $\xi_1$, the least studied in the literature, which is the coupling that generates
all three Lorentz-violating coefficients. 
And such coefficients are responsible for triggering the spontaneous Lorentz
violation in the SME, which is the context adopted in this article. 

The phenomenology of the coupling $\xi_1$ was analyzed using the Bianchi I metric as the cosmological spacetime,
a spacetime that is not in agreement with the cosmological principle.
In that context, the cosmological one, the Lorentz-violating coupling produced a small deviation from the 
standard geometry in cosmology, namely the FLRW geometry, inducting then a 
faster acceleration in a given spatial direction
during a dark energy era. Therefore, a phenomenological aspect of the coupling constant $\xi_1$ 
was brought out in this article, and also a range of validity for that coupling constant was achieved.

Lastly, as a development of ideas from this article, the inflationary period could be studied
in a future work by assuming both the Kalb--Ramond field within the Lorentz symmetry breaking context 
 (something that is missing in inflation models) and the Bianchi I cosmology in order to reproduce an anisotropic 
 inflationary model.

\begin{acknowledgments}
RVM thanks Fundação Cearense de Apoio ao Desenvolvimento
Científico e Tecnológico (FUNCAP), Coordenação de Aperfeiçoamento de Pessoal de Nível Superior (CAPES), 
and Conselho Nacional de Desenvolvimento Científico e Tecnológico (CNPq, Grant no 307556/2018-2) 
for the financial support.
\end{acknowledgments}

\end{document}